\documentclass[aps,preprint,showpacs,amsmath,amssymb,prd,nofootinbib]{revtex4}
\usepackage{amssymb,graphics,graphicx,epstopdf,color}

\begin{document}

\title
{\Large \bf An $A_5$ Model of Four Lepton Generations}

\author{Chian-Shu~Chen$^{1,2}$\footnote{chianshu@phys.sinica.edu.tw}, Thomas W. Kephart$^{3}$\footnote{tom.kephart@gmail.com}, and Tzu-Chiang Yuan$^{2}$\footnote{tcyuan@phys.sinica.edu.tw}}
\affiliation{$^{1}$Physics Division, National Center for Theoretical Sciences, Hsinchu, Taiwan 300\\$^{2}$Institute of Physics, Academia Sinica, Taipei, Taiwan 115\\$^{3}$Department of Physics and astronomy, Vanderbilt University, Nashville, Tennessee 37235, USA}

\date{\today}

\begin{abstract}
We study the lepton sector of a  four generations model  based on the discrete flavor group $A_5$. The best features of the three family $A_4$ model survive, including the tribimaximal pattern of three generation neutrino mixings. 
At leading order the three light neutrino mass relations
of $m_{\nu_1} = m_{\nu_3}$ and $m_{\nu_2} = 0$ are predicted.
The splitting of the neutrino masses can be naturally obtained as a result of the breaking of $A_{5}$ down to $A_4$ and a degenerate spectrum is preferred in our model.
 The electron mass is zero at tree level, but calculable through quantum corrections in our $A_5$ model.

\noindent
\end{abstract}

\maketitle

\section{Introduction}

Recently there has been considerable efforts to explore 
the possibility of a fourth generation of fermions in the 
standard model (SM). In particular these studies of fourth generation 
include its constraints by current electroweak precision 
data \cite{He:2001tp,Erler-Langacker,Chanowitz,PQ-Sher}, 
its interplay with Higgs physics \cite{Higgs4th}, 
CP violation effects in $B$-physics \cite{LHCb} and 
accessibility
of the fourth generation quarks~\cite{Holdom} and leptons~\cite{CRW}  at the Large Hadron Collider (LHC) .
For a recent review of fourth generation physics, see for example Ref.~\cite{overview}. 

In this paper, we will focus mainly on the fourth generation of leptons and comment
only briefly on inclusion of the extra quark family toward the end.
If we assume the flavor symmetry of the lepton sector is described correctly at tree level by an $A_4$ model \cite{Ma:2001dn, Altarelli:2005yp, Zee:2005ut, He:2006dk, King:2006np, Morisi:2007ft, Bazzocchi:2007na, Frampton:2008ci}, we then need to reconcile the prospects of a heavy fourth generation of leptons with the results known for the neutrinos in the first three generations. This requires that we extend the successful results of $A_4$ models, such as tribimaximal mixing (TBM), to the four generation model. The simplest approach is to demand a model where the $A_4$ model can be minimally embedded. The simplest extension of this type is based on  $A_5$ where surprisingly many new features arise, some of which will be immediately testable in the next round of neutrino experiments or at the LHC.

In the literature there has been numerous discussions of three generation lepton flavor symmetry using the $A_{5}$ group.
It is well known that the tribimaximal value of the solar mixing angle $\theta^{\rm TBM}_{12} = \tan^{-1}(\frac{1}{\sqrt{2}}) = 35.26\,^{\circ}$, which is very close to the experimental best-fit value of $34.43\,^{\circ}$.
In Refs.\cite{Datta:2003qg, Kajiyama:2007gx} the solar mixing angle is related to the golden ratio 
$\phi=(1+\sqrt{5})/2$~\cite{ratio} via $\theta_{12} = \tan^{-1}(\frac{1}{\phi}) = 31.72\,^{\circ}$ which is $2\sigma$ below the experimental best-fit value.  
The icosahedral group $I$, isomorphic to the permutation group $A_{5}$, has also been utilized in  lepton flavor symmetry in Refs.~\cite{Cummins:1988dr,shirai,Luhn:2007yr} (also see recent reviews in ~\cite{Ludl:2009ft,Ishimori:2010au}) where the golden ratio also shows up naturally. 
This suggests a possible connection to $A_5$ as shown in Ref.~\cite{Everett:2008et}.
Another interesting parameterization involving the golden ratio is ${\theta_{12}} = {\frac{\pi}{5}} = 
\cos^{-1} (\frac{\phi}{2})$ and $\theta_{\rm cabibbo} = \frac{\pi}{12}$ introduced in 
Ref.~\cite{Rodejohann:2008ir} for both the lepton and quark sectors. This might have connection with
the $A_5$ group as well.

In this paper we propose the  four generation lepton sector fit  into the context of an $A_{5}\times Z_{2}\times Z_{3}$ symmetry where $A_5$ subsequently breaks into $A_{4}$, with the goal of retaining the tribimaximal mixing as well as the lepton mass pattern (three hierarchical charged leptons and three tiny neutrino masses). In Section 2 we begin with a brief introduction of the discrete group $A_{5}$ (for more details see Appendix~A) and then utilize the structure to the products of its irreducible representations to construct a four generation lepton model. In Section 3 we work out the lepton mass spectrum and discuss how the tribimaximal mixing is embedded
in this model. Some of the rich flavor physics phenomenologies of  this model are also discussed. 
Finally, we conclude in Section 4 with some comments on collider physics implications. We also comment on the natural extension of our $A_5$ model to
the double icosahedral group $I'$ as the flavor symmetry to 
describe four generations of both quarks and leptons.

\section{The model}

There are five irreducible representations (irreps) of $A_{5}$: one singlet \textbf{1}, two triplets, $\textbf{3}$ and $\textbf{3}'$, one quartet $\textbf{4}$ and one quintet $\textbf{5}$. The character table and the multiplication rules of the five irreducible representations are given in Appendix~A. 

Our model is based on $A_{5}$ with an additional $Z_{2}\times Z_{3}$ symmetry to avoid unwanted terms in the Lagrangian. 
We assign the four generations of left-handed lepton doublets to $A_{5}\times Z_{2}\times Z_{3}$ as follows:
\begin{eqnarray}\label{lepdbt}
\underbrace{\left(\begin{array}{c}\nu_{4} \\ l_{4} \end{array}\right)_{L} \;\;
\left(\begin{array}{c}\nu_{3} \\ l_{3} \end{array}\right)_{L} \;\;
\left(\begin{array}{c}\nu_{2} \\ l_{2} \end{array}\right)_{L} \;\;
\left(\begin{array}{c}\nu_{1} \\ l_{1} \end{array}\right)_{L}}_{L_{L4} (\textbf{4}, +1,\omega)} \;\;, 
\end{eqnarray}
where $\left(\begin{array}{c}\nu_{i} \\ l_{i} \end{array}\right)_{L}$ with $i=1,2,3,4$ represents the left-handed lepton doublets of the $i^{\text{th}}$ family and $\omega = e^{i2\pi/3}$ is the cubic root of unity. 
For the right-handed sector we introduce the charged lepton fields 
\begin{eqnarray}\label{leprt}
l_{R5} (\textbf{5}, -1, +1) \quad, \quad  l^{c}_{R3} (\textbf{3}, -1, +1) \quad {\rm and} \quad l^{(1)}_{R1} (\textbf{1}, -1, +1) \quad, \quad l^{(2)}_{R1} (\textbf{1}, -1, +1) 
\end{eqnarray}
which transform as a $\bf 5$, a $\bf 3$, and two trivial singlets $\bf 1$ under $A_{5}$ symmetry respectively. Here $c$ stands for charge conjugation. The right-handed neutrinos are assigned as one quintet $\textbf{5}$ and a trivial singlet  
\begin{eqnarray}\label{nu}
N_{R5} (\textbf{5}, +1, +1) \quad {\rm and} \quad N^{(1)}_{R} (\textbf{1}, +1, +1) \;\;. 
\end{eqnarray}
We have implicitly embedded  the right-handed charged leptons $\mu_{R}$, $e_{R}$
in the 5 dimensional irreducible representation (irrep) of $A_{5}$. 

For the Higgs sector we introduce three $A_{5}$ quartet and one $A_{5}$ triplet scalars. The first quartet $S_{4}$ is a gauge singlet which transform as  $(\textbf{4},+1,+1)$ under  $A_{5}\times Z_{2}\times Z_{3}$. The other two quartets are electroweak doublets with $Z_2 \times Z_3$ charges and are written as $H_{4} (\textbf{4}, +1,\omega^2)$ and $H'_{4} (\textbf{4}, -1,\omega^2)$. The $A_5$ triplet  $\Phi_{3}(\textbf{3}, +1, \omega^2)$ is also an electroweak doublet and has nontrivial $Z_3$ charge. The most general renormalizable Lagrangian of the 
Yukawa couplings among these fields and Majorana mass terms for the neutrinos 
that are invariant under both the standard model gauge group 
and the $A_{5}\times Z_{2}\times Z_{3}$ discrete flavor symmetry is
\begin{eqnarray}\label{A5Yukawa}
\emph{L}_{\rm {Yukawa}} &=& \frac{1}{2}M_{1}N^{(1)}_{R}N^{(1)}_{R} + \frac{1}{2}M_5N_{R5}N_{R5} + Y_{S1}(S_{4}N_{R5}N_{R5}) + Y_{S2}(S_{4}(l^{-}_{R3})^{c}l^{-}_{R5}) \nonumber \\
&+& Y_{1}(L_{L4}N^{(1)}_{R}H_{4}) + Y_{2}(L_{L4}N_{R5}H_{4}) + Y_{3}(L_{L4}N_{R5}\Phi_{3}) \nonumber \\ 
&+& Y_{4}(L_{L4}l_{R5}H'_{4}) + Y_{5}(L_{L4}l^{(1)}_{R1}H'_{4}) + Y_{6}(L_{L4}l^{(2)}_{R1} H'_{4}) + \rm {H.c.} 
\end{eqnarray}
Suppose the scalar $S_{4}$ develops a VEV $\langle S_{4} \rangle = (V_{S}, 0, 0, 0)$. It will break the discrete group $A_{5}$ into $A_{4}$, causing the irreps of $A_{5}$ to decompose as 
\begin{eqnarray}
&A_{5}& ~\rightarrow  A_{4}\nonumber \\
&\textbf{1}& \rightarrow \textbf{1} \nonumber \\
&\textbf{3}& \rightarrow \textbf{3} \nonumber \\
&\textbf{3}'& \rightarrow \textbf{3} \nonumber \\
&\textbf{4}& \rightarrow \textbf{1} + \textbf{3} \nonumber \\
&\textbf{5}& \rightarrow \textbf{1}' + \textbf{1}'' + \textbf{3} \nonumber
\end{eqnarray}
Hence the lepton fields are decomposed as 
$$L_{L_{4}} \rightarrow L_{L1} (\textbf{1}, +1,\omega) + L_{L3} (\textbf{3},+1,\omega) \;\;,$$
$$l_{R5} \rightarrow l_{R3} (\textbf{3}, -1, +1) + l_{R1'} (\textbf{1}', -1, +1) + l_{R1''} (\textbf{1}'',-1, +1) \;\;,$$ 
and 
$$N_{R5} \rightarrow N_{R3} (\textbf{3},+1, +1) + N^{(2)}_{R} (\textbf{1}',+1, +1) + N^{(3)}_{R} (\textbf{1}'',+1, +1) \;\;.$$
Here we assume the right-handed $A_{4}$ triplet $l_{R3}$ is combined with the lepton $(l)^{c}_{R3}$ shown in Eq.~(\ref{leprt}) to form a vector field $l_{L+R,3}$ so as to insure gauge anomaly cancellation. Similarly, the Higgs scalars decompose  as 
$$S_{4} \rightarrow S_{1}(\textbf{1}, +1, +1) + S_{3}(\textbf{3}, +1, +1) \;\;,$$ 
$$H_{4} \rightarrow H_{1} (\textbf{1}, +1, \omega^2) + H_{3} (\textbf{3}, +1, \omega^2) \;\;,$$  
$$H'_{4} \rightarrow H'_{1} (\textbf{1}, -1, \omega^2) + H'_{3} (\textbf{3}, -1, \omega^2) \;\;,$$
and
$$\Phi_{3} \rightarrow \Phi_{3}(\textbf{3}, +1, \omega^2) \;\;.$$  
 After this stage of $A_{5} \rightarrow A_{4}$ breaking via $\langle S_{4} \rangle$, the Yukawa Lagrangian can be written as
\begin{eqnarray}\label{A4potential}
L^{A_{4}}_{Yukawa} &=& \frac{1}{2}M_{1}N^{(1)}_{R}N^{(1)}_{R} + \frac{1}{2}M_{5}(N^{(2)}_{R}N^{(3)}_{R} + N_{R3}N_{R3}) \nonumber \\
&+& Y_{S1} \left[(V_{S} + S_{1})(N^{(2)}_{R}N^{(3)}_{R} + N_{R3}N_{R3}) + S_{3}(N_{R3}N_{R3}) \right] \nonumber \\
&+& Y_{S2}\left[(V_{S} + S_{1})(l_{R3})^{c}l_{R3} + S_{3}(l_{R3})^{c}(l_{R3} + l_{R1'} + l_{R1''}) \right] \nonumber \\
&+& Y_{1}\left[L_{L1}N^{(1)}_{R}H_{1} + L_{L3}N_{R}^{(1)}H_{3} \right] \nonumber \\
&+& Y_{2}\left[L_{L1}N_{R3}H_{3} + L_{L3}N_{R3}(H_{1} + H_{3}) + L_{L3}(N^{(2)}_{R} + N^{(3)}_{R})H_{3}\right] \nonumber \\
&+& Y_{3}\left[L_{L1}N_{R3}\Phi_{3} + L_{L3}N_{R3}\Phi_{3} + L_{L3}(N^{(2)}_{R} + N^{(3)}_{R})\Phi_{3}\right] \nonumber \\
&+& Y_{4}\left[L_{L1}l_{R3}H'_{3} + L_{L3}l_{R3}(H'_{1} + H'_{3}) + L_{L3}(l_{R1'} + l_{R1''})H'_{3}\right] \nonumber \\
&+& Y_{5}\left[L_{L1}l^{(1)}_{R1}H'_{1} + L_{L3}l^{(1)}_{R1}H'_{3} \right] + Y_{6}\left[L_{L1}l^{(2)}_{R1}H'_{1} + L_{L3}l^{(2)}_{R1}H'_{3}\right] + {\rm H.c.}
\end{eqnarray} 
For convenience, the multiplication rules for $A_4$ 
are summarized in Appendix~B.

\section{Phenomenology}

We begin with the mass spectrum of the leptons.

\subsection{Charged Lepton Masses} 

\begin{figure}[t]
  \centering
\includegraphics[width=0.3\textwidth]{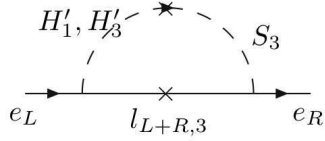}
  \caption{A 1-loop diagram which gives the electron a finite small mass through new vector leptons.}
  \label{fig:e-mass}
\end{figure}

After the subsequent breaking of the $A_{4}$ and SM 
gauge symmetries due to the VEVs of the 
scalar fields $S_{3}, H_{1}, H_{3}, H'_{1}$ and $H'_{3}$,
the charged lepton mass terms are given by
\begin{eqnarray}
L_{l} &=& Y_{S2}\left[(V_{S} + \langle S_{1}\rangle)(l_{R3})^{c}l_{R3} + \langle S_{3}\rangle(l_{R3})^{c}(l_{R3} + l_{R1'} + l_{R1''}) \right] \nonumber \\
&+& Y_{4}\left[L_{L1}l_{R3}\langle H'_{3}\rangle + L_{L3}l_{R3}(\langle H'_{1} \rangle + \langle H'_{3}\rangle) + L_{L3}(l_{R1'} + l_{R1''})\langle H'_{3}\rangle \right]\nonumber \\
&+& Y_{5}\left[L_{L1}l^{(1)}_{R1}\langle H'_{1}\rangle + L_{L3}l^{(1)}_{R1}\langle H'_{3}\rangle\right] + Y_{6}\left[L_{L1}l^{(2)}_{R1}\langle H'_{1}\rangle + L_{L3}l^{(2)}_{R1}\langle H'_{3}\rangle\right].
\end{eqnarray}
We take the VEVs of $H'_{3}$ and $H'_{1}$ to be
\begin{eqnarray}
\langle H'_{3} \rangle = (V'_{3_1}, V'_{3_2}, V'_{3_3}) \quad {\rm and} \quad \langle H'_{1} \rangle = V'_{1} \;\;,
\end{eqnarray}
which yields the $7\times7$ charged lepton mass matrix of the form 
\begin{eqnarray}
M_{l} = \left(\begin{array}{cccc|ccc}Y_{5}V'_{1} & Y_{6}V'_{1} & 0 & 0 & Y_{4}V'_{3_1} & Y_{4}V'_{3_2} & Y_{4}V'_{3_3} \\Y_{5}V'_{3_1} & Y_{6}V'_{3_1} & Y_{4}V'_{3_1} & Y_{4}V'_{3_1} & Y_{4}V'_{1} & Y_{4}V'_{3_3} & Y_{4}V'_{3_2} \\Y_{5}V'_{3_2} & Y_{6}V'_{3_2} & \omega Y_{4}V'_{3_2} & \omega^2 Y_{4}V'_{3_2} & Y_{4}V'_{3_3} & Y_{4}V'_{1} & Y_{4}V'_{3_1} \\Y_{5}V'_{3_3} & Y_{6}V'_{3_3} & \omega^2 Y_{4}V'_{3_3} & \omega Y_{4}V'_{3_3} & Y_{4}V'_{3_2} & Y_{4}V'_{3_1} & Y_{4}V'_{1} \\\hline 0 & 0 & Y_{S2}\langle S_{3} \rangle_{1} & Y_{S2}\langle S_{3} \rangle_{1} & Y_{S2}(V_{S} + \langle S_{1} \rangle) & Y_{S2}\langle S_{3} \rangle_{3} & Y_{S2}\langle S_{3} \rangle_{2} \\ 0 & 0 & \omega Y_{S2}\langle S_{3} \rangle_{2} & \omega^2Y_{S2}\langle S_{3} \rangle_{2} & Y_{S2}\langle S_{3} \rangle_{3} & Y_{S2}(V_{S} + \langle S_{1} \rangle) & Y_{S2}\langle S_{3} \rangle_{1} \\
0 & 0 & \omega^2Y_{S2}\langle S_{3}\rangle_{3} & \omega Y_{S2}\langle S_{3}\rangle_{3} & Y_{S2}\langle S_{3} \rangle_{2} & Y_{S2}\langle S_{3} \rangle_{1} & Y_{S2}(V_{S} + \langle S_{1} \rangle) \end{array}\right)\nonumber \\
\end{eqnarray}
written in the left-handed and right-handed charged leptons bases given by
$(L_{L1}, L_{L3}= (L_{L3_1}, L_{L3_2}, L_{L3_3}),l^c_{R3}=(l^c_{R3_1},l^c_{R3_{2}},l^c_{R3_{3}}))$ 
and $(l^{(1)}_{R1}, l^{(2)}_{R1}, l_{R1'}, l_{R1''},l_{R3_{1}},l_{R3_{2}},l_{R3_{3}})^T$ respectively. 
Note that the first two columns of the mass matrix are proportional to each other, hence the determinant of this matrix is zero, Det$(M_{l}) = 0$, which implies it has a zero eigenvalue to be 
identified as the electron mass. We thus predicts the electron is massless at tree level in our model. This is because the two $A_{5}$ singlet right-handed charged leptons (see Eq.~(\ref{leprt})) have the same  Yukawa couplings structure due to the $A_5$ multiplication rules and this result is independent of the values of  VEVs. The construction provides another example of electron-muon universality proposed in Ref~\cite{Barr:1976bk}, where a class of models was devised to evaluate the small ratio  $m_{e}/m_{\mu}$.
Under the assumption that the  $A_{5}$ breaking scale is much higher than those for the 
breaking of $A_{4}$ and the SM gauge symmetries, 
namely $V_{S} \gg \langle S_{1}\rangle, \langle S_{3}\rangle, \langle H'_{1}\rangle$, and $\langle H'_{3}\rangle$, we can treat the vector field $l_{L+R,3}$ as being decoupled from the four chiral lepton generations, while their mixings can be ignored at leading order. 
If we set $V'_{3_1} = V'_{3_2} = V'_{3_3} = V'$ for simplicity, 
then we obtain the following charged lepton masses for the four generations
\begin{eqnarray}\label{chargemass}
m_{e} &=& 0 \quad {,} \quad m_{\mu} = \sqrt{3}Y_{4}V'  \;\;,\nonumber \\
m_{\tau} &=& \frac{1}{2}\Big[(Y_{5}V'_{1} + (Y_{6} - Y_{4})V') - \sqrt{(Y_{5}V'_{1} + (Y_{6} - Y_{4})V')^2 + 4Y_{4}(Y_{5}V'_{1} + 3Y_{6}V')V'} \Big] \;\;, \nonumber 
\end{eqnarray}
and
\begin{eqnarray}
m_{\tau'} &=& \frac{1}{2}\Big[(Y_{5}V'_{1} + (Y_{6} - Y_{4})V') + \sqrt{(Y_{5}V'_{1} + (Y_{6} - Y_{4})V')^2 + 4Y_{4}(Y_{5}V'_{1} + 3Y_{6}V')V'} \Big] \;\;.
\label{tauprimemass}
\end{eqnarray}  
A $95\%$ C.L. lower mass limit for the heavy charged leptons is set around $100$ GeV~\cite{Achard:2001qw} which can be used to give bounds on the unknown parameters in Eq.(\ref{tauprimemass}). 
A finite electron mass can be generated via a 1-loop diagram through the intermediate vector fermions as illustrated in Fig.~\ref{fig:e-mass}. The result is similar to the many earlier attempts to calculate electron mass~\cite{Georgi:1972hy, Barr:1976bk,Barr:1978rv,Balakrishna:1988xg} or fermion spectrum~\cite{Ibanez:1982xg,Barr:2007ma}. The electron mass is estimated to be
\begin{eqnarray}
m_{e} \sim \frac{Y^2_{4}}{16\pi^2}\frac{m^2_{H}}{m_{L+R,3}} \approx \frac{Y^2_{4}}{16\pi^2Y_{S2}}\frac{m^2_{H}}{V_{S}} \;\;,
\end{eqnarray}  
where we ignore both mixings in lepton and scalar sectors and
$m_{H}$ represents the common mass scale of scalars for simplicity. 
By using Eq.~(\ref{chargemass}) we 
have the relation $m_{e}/m_{\mu} \approx \frac{Y_{4}m^2_{H}}{16\pi^2Y_{S2}V_{S}V'} \approx \frac{\alpha}{\pi}$. We should point out that the result depends sensitively on several parameters as 
in many previous works~\cite{Georgi:1972hy, Barr:1976bk,Barr:1978rv,Balakrishna:1988xg,Ibanez:1982xg}; however, a recent study in Refs.~\cite{Barr:2007ma} indicates that the parameters could be reduced considerably.  

Note that the charged lepton are given masses  by the VEV $\langle H'_{4} \rangle$ which is not directly related to the neutrino masses, so we can calculate the mass eigenstates of charged leptons by using the biunitary transformations to diagonalize the mass matrix $M_{l}$. The rotation matrix of left-handed fields can be absorbed into the redefinition of Higgses $H'_{1}$ and $H'_{3}$, while the rotation of the right-handed charged leptons will not affect the tribimaximal mixings for neutrinos at leading order.

\subsection{Neutrino Masses}

We next turn to the neutrino masses. 
The full neutrino mass matrix in the model 
is a $10\times10$ matrix, written as 
\begin{eqnarray}
M_{\nu} = \left(\begin{array}{c|c}M'_{\nu_L}(4\times4)=0 & M_{D}(4\times6) \\\hline M^T_{D}(6\times4) & M_{N_{R}}(6\times6)\end{array}\right) \;\; .
\end{eqnarray}

From Eq.~(\ref{A4potential}) we obtain the $6\times6$ right-handed neutrino mass matrix\\
 in the  $(N^{(1)}_{R}, N^{(2)}_{R}, N^{(3)}_{R}, N_{R3_{1}}, N_{R3_{2}}, N_{R3_{3}})$ basis  
\begin{eqnarray}\label{MR}
M_{N_{R}} = \left(\begin{array}{cccccc}M_{1} & 0 & 0 & 0 & 0 & 0 \\0 & 0 & M_{2} & 0 & 0 & 0 \\0 & M_{2} & 0 & 0 & 0 & 0 \\0 & 0 & 0 & 2M_{2} & 0 & 0 \\0 & 0 & 0 & 0 & 2M_{2} & 0 \\0 & 0 & 0 & 0 & 0 & 2M_{2} \end{array}\right)  \;\;,
\end{eqnarray}
where we define $M_{2} = \frac{1}{2}M_{5} + Y_{S1}V_{S}$. 
Since the $A_{5}$ breaking scale $V_{S}$ is assumed to be higher than the subsequent breaking scales of the $A_{4}$ and SM gauge symmetries, we ignore the effects of $\langle S_{3} \rangle$ in the following discussion of the derivation of  tribimaximal mixings and we will treat these effects as perturbations when we calculate the neutrino mass spectrum.

The Dirac mass terms of the neutrino sector are given by 
\begin{eqnarray}
L_{Dirac} &=& Y_{1}\left[L_{L1}N^{(1)}_{R}H_{1} + L_{L3}N_{R}^{(1)}H_{3} \right] + Y_{2}L_{L3}N_{R3}H_{1} + L_{L1}N_{R3}\left[Y_{2}H_{3} + Y_{3}\Phi_{3}\right] \nonumber \\
&+& L_{L3}N_{R3}\left[Y_{2}H_{3} + Y_{3}\Phi_{3}\right] + L_{L3}(N^{(2)}_{R} + N^{(3)}_{R})\left[Y_{2}H_{3} + Y_{3}\Phi_{3}\right] \;\;.
\end{eqnarray}
We see that the linear combination of the two $A_{4}$ triplet fields, $H_{3}$ and $\Phi_{3}$, contribute to the Dirac  neutrino masses. One can decouple the fourth generation neutrino from the three light generations and satisfy the conditions of tribimaximal mixings by assuming the VEVs of $H_{3}$ are relatively small compare to the VEVs of $\Phi_{3}$ as we will demonstrate below. Hence the Dirac matrix of the neutrino sector $M_{D}$ is given by
\begin{eqnarray}
&&\nu_{L}M_{D}N^T_{R} = \nonumber \\
&&\left(\begin{array}{cccc}\nu_{L1} & \nu_{L3_1} & \nu_{L3_2} & \nu_{L3_3}\end{array}\right)
\left(\begin{array}{cccccc}Y_{1}v_{H_1} & 0 & 0 & Y_{3}v_{\phi_1} & Y_{3}v_{\phi_2} & Y_{3}v_{\phi_3} \\ 0 & Y_{3}v_{\phi_1} & Y_{3}v_{\phi_1} & Y_{2}v_{H_1} & Y_{3}v_{\phi_3} & Y_{3}v_{\phi_2} \\ 0 & \omega Y_{3}v_{\phi_2} & \omega^2Y_{3}v_{\phi_2} & Y_{3}v_{\phi_3} & Y_{2}v_{H_1} & Y_{3}v_{\phi_1} \\ 0 & \omega^2Y_{3}v_{\phi_3} & \omega Y_{3}v_{\phi_3} & Y_{3}v_{\phi_2} & Y_{3}v_{\phi_1} & Y_{2}v_{H_1}\end{array}\right)\left(\begin{array}{c}N^{(1)}_{R} \\N^{(2)}_{R} \\N^{(3)}_{R} \\N_{R3_{1}} \\N_{R3_{2}} \\ N_{R3_{3}}\end{array}\right) 
\nonumber  \;\;,\\
\end{eqnarray} 
where we set the VEVs $\langle \Phi_{3}\rangle = (v_{\phi_1}, v_{\phi_2}, v_{\phi_3})$ and $\langle H_{1}\rangle = v_{H_1}$. 

The left-handed Majorana mass matrix $M_{\nu_{L}}$ is obtained by the seesaw mechanism, 
\begin{eqnarray}
M_{\nu_{L}} = M_{D}M^{-1}_{N_{R}}M^{T}_{D} \;\;, 
\end{eqnarray}
thus the ten components of the $4\times4$ symmetric matrix can be written as 
\begin{eqnarray}
M_{\nu_L(\nu_{L1}, \nu_{L1})} &=& \frac{Y^2_{1}v^2_{H_1}}{M_1} + \frac{Y^2_{3}(v^2_{\phi_1} + v^2_{\phi_2} + v^2_{\phi_3})}{2M_2} \;\; , \nonumber \\
M_{\nu_L(\nu_{L1}, \nu_{L3_1})} &=& \frac{Y_{2}Y_{3}v_{\phi_1}v_{H_1} + 2Y^2_{3}v_{\phi_2}v_{\phi_3}}{2M_2} \;\; ,  \nonumber \\
M_{\nu_L(\nu_{L1}, \nu_{L3_2})} &=& \frac{Y_{2}Y_{3}v_{\phi_2}v_{H_1} + 2Y^2_{3}v_{\phi_1}v_{\phi_3}}{2M_2} \;\; ,  \nonumber \\
M_{\nu_L(\nu_{L1}, \nu_{L3_3})} &=& \frac{Y_{2}Y_{3}v_{\phi_3}v_{H_1} + 2Y^2_{3}v_{\phi_1}v_{\phi_2}}{2M_2} \;\; ,  \nonumber \\
M_{\nu_L(\nu_{L3_1}, \nu_{L3_1})} &=& \frac{Y^2_{2}v^2_{H_1} + Y^2_{3}(4v^2_{\phi_1} + v^2_{\phi_2} + v^2_{\phi_3})}{2M_2}  \;\; ,  \\ 
M_{\nu_L(\nu_{L3_1}, \nu_{L3_2})} &=& \frac{2Y_{2}Y_{3}v_{H_1}v_{\phi_3} - Y^2_{3}v_{\phi_1}v_{\phi_2}}{2M_2}  \;\; , \nonumber \\
M_{\nu_L(\nu_{L3_1}, \nu_{L3_3})} &=& \frac{2Y_{2}Y_{3}v_{H_1}v_{\phi_2} - Y^2_{3}v_{\phi_1}v_{\phi_3}}{2M_2}  \;\; , \nonumber \\
M_{\nu_L(\nu_{L3_2}, \nu_{L3_2})} &=& \frac{Y^2_{2}v^2_{H_1} + Y^2_{3}(v^2_{\phi_1} + 4v^2_{\phi_2} + v^2_{\phi_3})}{2M_2} \;\; ,  \nonumber \\
M_{\nu_L(\nu_{L3_2}, \nu_{L3_3})} &=& \frac{2Y_{2}Y_{3}v_{H_1}v_{\phi_1} - Y^2_{3}v_{\phi_2}v_{\phi_3}}{2M_2}  \;\; , \nonumber \\
M_{\nu_L(\nu_{L3_3}, \nu_{L3_3})} &=& \frac{Y^2_{2}v^2_{H_1} + Y^2_{3}(v^2_{\phi_1} + v^2_{\phi_2} + 4v^2_{\phi_3})}{2M_2} \;\; . \nonumber  
\end{eqnarray}
We can diagonalize the $4\times4$ symmetric mass matrix $M_{\nu_L}$  by writing it in the form
\begin{eqnarray}
M_{\nu_L} = \left(\begin{array}{cccc}A & B & C & D \\B & E & F & G \\C & F & H & I \\D & G & I & J\end{array}\right)
\end{eqnarray} 
in the $(\nu_{\tau'}, \nu_{\tau}, \nu_{\mu}, \nu_{e})$ basis,
and then using the unitary transformation, 
\begin{eqnarray}
M_{diag} = U^{\rm 4g}_{TBM}M_{\nu_L}U^{4g^T}_{TBM} \;\;,
\end{eqnarray}
 where the 4 generations mixing matrix in the tribimaximal mixing limit can be expressed as  
\begin{eqnarray}
U^{4g}_{TBM} = \left(\begin{array}{cccc} 1 & 0 & 0 & 0 \\ 0 & \frac{1}{\sqrt{2}} & -\frac{1}{\sqrt{2}} & 0 \\0 & \sqrt{\frac{1}{3}} & \sqrt{\frac{1}{3}} & \sqrt{\frac{1}{3}} \\0 & -\sqrt{\frac{1}{6}} & -\sqrt{\frac{1}{6}} & \sqrt{\frac{2}{3}} \end{array}\right).
\end{eqnarray}
There are six conditions corresponding to the vanishing off diagonal terms, viz. 
\begin{eqnarray}
\label{con1}
&&E = H \quad {\rm ,} \quad G = I \;\; ,\\
\label{con2}
&&G + J = E + F \;\; , \\
&&B = C = D = 0 \;\; .
\label{con3}
\end{eqnarray}
Eq.~(\ref{con1}) implies $v_{\phi_1} = v_{\phi_2}$ and the solution for Eq.~(\ref{con2}) is
\begin{eqnarray}
Y_{3}(v_{\phi_1} - v_{\phi_3})\left[Y_{3}(2v_{\phi_1} + 3v_{\phi_3}) -2Y_{2}v_{H_1}\right]
& = & 0 \; . 
\end{eqnarray}
Together with the conditions from Eq.~(\ref{con3}) we have only one consistent solution, $v_{\phi_1} = v_{\phi_2}$ = $v_{\phi_3} \equiv v$. Thus we write 
\begin{eqnarray}\label{vacuum} 
\langle \Phi_{3} \rangle= (v,v,v) 
\end{eqnarray} 
and the constraint from Eq.~(\ref{con3}) becomes
\begin{eqnarray}
v_{H_1} = -\frac{2Y_{3}}{Y_{2}}v \;\;.
\end{eqnarray}
This leads to the four neutrino masses  
\begin{eqnarray}
m_{\nu_4} &=& \frac{Y^2_{1}v^2_{H_1}}{M_1} + \frac{3Y^2_{3}v^2}{2M_{2}} \;\; , \nonumber \\
m_{\nu_3} &=& \frac{15Y^2_{3}v^2}{2M_{2}} \;\; , \nonumber \\
m_{\nu_2} &=& 0 \;\; , \nonumber \\
m_{\nu_1} &=& m_{\nu_3} \; \; .
\label{neutrinosmass0}
\end{eqnarray}
The parameters in the first term of $m_{\nu_4}$ 
all differ from those in the expressions for the other three SM neutrinos, hence they can be used
to easily satisfy the experimental bound $m_{\nu_4} > M_{Z}/2$. 
At leading order the three light neutrino masses are $m_{\nu_1} = m_{\nu_3}$ and $m_{\nu_2} = 0$, which are phenomenological unacceptable. However, the inclusion of perturbative effects on the masses due to the heavy Majorana neutrinos through the VEVs of $S_{3}$ will correct the above three light neutrinos masses and allow a good fit to the oscillation data. We note that this perturbative effect is from the assumed $A_5$ symmetry and its subsequent spontaneous break down to $A_4$. The situation is similar to the proposal in Ref.~\cite{Ma:2004zv}, in which the neutrino masses can be split  by the small VEVs of several heavy Higgs triplets which are assigned to different representations of the $A_{4}$ group. 
Therefore, taking the VEVs of $S_{3}$, to be $\langle S_{3} \rangle = (\delta_1,\delta_2,\delta_3)$, the perturbations to the right-handed Majorana mass matrix, as shown in Eq.~(\ref{MR}) are
\begin{eqnarray}\label{MRp}
M_{N_{R}} = \left(\begin{array}{cccccc}M_{1} & 0 & 0 & 0 & 0 & 0 \\0 & 0 & M_{2} & 0 & 0 & 0 \\0 & M_{2} & 0 & 0 & 0 & 0 \\0 & 0 & 0 & 2M_{2} & Y_{S1}\delta_3 & Y_{S1}\delta_2 \\0 & 0 & 0 & Y_{S1}\delta_3 & 2M_{2} & Y_{S1}\delta_1 \\0 & 0 & 0 & Y_{S1}\delta_2 & Y_{S1}\delta_1 & 2M_{2} \end{array}\right) \;\; .
\end{eqnarray}
The three light neutrino masses become
\begin{eqnarray}
m_{\nu_3} &\approx& \frac{6Y^2_{3}v^2}{M_{2}} + \frac{3}{2M_{2} + Y_{S1}\delta_{1}}Y^2_{3}v^2  \;\; , \nonumber \\
m_{\nu_2} &\approx& -\frac{3Y^2_{3}v^2}{2M_{2}} + \frac{2Y^2_{3}v^2}{2M_{2} - Y_{S1}\delta_{1}} + \frac{(2M_{2} + Y_{S1}\delta_{1})Y^2_{3}v^2}{M_{2}(2M_{2} - Y_{S1}\delta_{1})}  \;\; , \nonumber \\
&& - \frac{(2M_{2} + Y_{S1}\delta_{1})Y^2_{3}v^2}{2M^2_{2}} + \frac{(2M_{2} + Y_{S1}\delta_{1})^2Y^2_{3}v^2}{2M_{2}(4M^2_{2} - Y^2_{S1}\delta^2_{1})}  \;\; , \nonumber \\
m_{\nu_1} &\approx& \frac{9Y^2_{3}v^2}{2M_{2}} + \frac{2Y^2_{3}v^2}{2M_{2} - Y_{S1}\delta_{1}} + \frac{(2M_{2} + Y_{S1}\delta_{1})Y^2_{3}v^2}{(2M_{2} - Y_{S1}\delta_{1})M_{2}}  \;\; , \nonumber \\
&& + \frac{(2M_{2} + Y_{S1}\delta_{1})Y^2_{3}v^2}{4M^2_{2}} + \frac{(2M_{2} + Y_{S1}\delta_{1})^2Y^2_{3}v^2}{2M_{2}(4M^2_{2} - Y^2_{S1}\delta^2_{1})}  \;\; , 
\end{eqnarray}
for $\delta_{1} \neq 0$ and $\delta_{2} = \delta_{3} = 0$. Similarly, for the cases of $(0, \delta_{2}, 0)$ and $(0,0,\delta_{3})$ we have  
\begin{eqnarray}
m_{\nu_3} &\approx& \frac{9Y^2_{3}v^2}{2M_{2}} + \left[\frac{6M_{2}}{4M^2_{2} - Y^2_{S1}\delta_{2}^2} + \frac{3}{2M_{2} - Y_{S1}\delta_{2}}\right]Y^2_{3}v^2  \;\; ,  \nonumber \\
m_{\nu_2} &\approx& -\frac{3Y^2_{3}v^2}{2M_{2}} + \frac{2Y^2_{3}v^2}{2M_{2} - Y_{S1}\delta_{2}} + \frac{(2M_{2} + Y_{S1}\delta_{2})Y^2_{3}v^2}{M_{2}(2M_{2} - Y_{S1}\delta_{2})}  \;\; , \nonumber \\
&& - \frac{(2M_{2} + Y_{S1}\delta_{2})Y^2_{3}v^2}{2M^2_{2}} + \frac{(2M_{2} + Y_{S1}\delta_{2})^2Y^2_{3}v^2}{2M_{2}(4M^2_{2} - Y^2_{S1}\delta^2_{2})}  \;\; ,  \nonumber \\
m_{\nu_1} &\approx& \frac{9Y^2_{3}v^2}{2M_{2}} + \frac{2Y^2_{3}v^2}{2M_{2} - Y_{S1}\delta_{2}} + \frac{(2M_{2} + Y_{S1}\delta_{2})Y^2_{3}v^2}{M_{2}(2M_{2} - Y_{S1}\delta_{2})} + \frac{(2M_{2} + Y_{S1}\delta_{2})^2Y^2_{3}v^2}{2M_{2}(4M^2_{2} - Y^2_{S1}\delta^2_{2})}  \;\; , 
\end{eqnarray}
and 
\begin{eqnarray}\label{numassp3}
m_{\nu_{3}} &\approx& \frac{3Y^2_{3}v^2}{M_{2}} + \frac{9Y^2_{3}v^2}{2M_{2} - Y_{S1}\delta_{3}} \;\; ,  \nonumber \\
m_{\nu_{2}} &\approx& -\frac{3Y^2_{3}v^2}{2M_{2}} + \frac{2Y^2_{3}v^2}{2M_{2} + Y_{S1}\delta_{3}} + \frac{Y^2_{3}v^2(2M_{2} + Y_{S1}\delta_{3})^2}{8M^3_{2}} \;\; ,  \nonumber \\
m_{\nu_{1}} &\approx& \frac{6Y^2_{3}v^2}{M_{2}} + \frac{Y^2_{3}v^2}{2M_{2} + Y_{S1}\delta_{3}} + \frac{3Y^2_{3}v^2(2M_{2} + Y_{S1}\delta_{3})}{4M^2_{2}} + \frac{Y^2_{3}v^2(2M_{2} + Y_{S1}\delta_{3})^2}{8M^3_{2}}  \;\; , 
\end{eqnarray}
respectively. 
Note that the above three sets of equations reduce to Eq.(\ref{neutrinosmass0}) when 
$\delta_i \rightarrow 0$ as they should. In these three cases we have the decoupling fourth neutrino mass which should be required to satisfy  
\begin{eqnarray}
m_{\nu_4} &=& \frac{Y^2_{1}v^2_{H_1}}{M_1} + (\frac{2}{2M_{2} + Y_{S1}\delta_{i}} + \frac{1}{2M_{2}})Y^2_{3}v^2 > M_{Z}/2 \;\; .
\end{eqnarray} 
By assuming the new physics to be around $\sim 1$ TeV, and taking  $v_{H_1} = 220$ GeV, $v = 10$ GeV, $Y_{S1}\delta_{i=1,2,3} = 400$ GeV, $M_{1} = 500$ GeV, $M_{2} = 10^8$ GeV, $Y_{1} = 1$, and $Y_{3} = 10^{-3}$, we have
\begin{eqnarray}\label{nu-mass}
m_{\nu_3} &\approx& 0.75\times10^{-2}~{\rm eV} \;\;, \quad m_{\nu_{2}} \approx 5.4\times10^{-6}~{\rm eV} \;\;, \quad m_{\nu_{1}} \approx 0.35\times10^{-2}~{\rm eV}\; \; ; 
\nonumber \\
m_{\nu_{3}} &\approx& 0.75\times10^{-2}~{\rm eV} \;\;, \quad m_{\nu_{2}} \approx 6.0\times10^{-9}~{\rm eV} \;\;,  \quad m_{\nu_{1}} \approx 0.70\times10^{-2}~{\rm eV} \;\; ; 
\nonumber \\
m_{\nu_{3}} &\approx& 0.75\times10^{-2}~{\rm eV} \;\;, \quad m_{\nu_{2}} \approx 6.0\times10^{-15}~{\rm eV} \;\;, \quad m_{\nu_{1}} \approx 0.85\times10^{-2}~{\rm eV} \;\; ;  
\end{eqnarray}
and 
\begin{eqnarray}
m_{\nu_{4}} \approx 96.8~{\rm GeV}  \;\; , 
\end{eqnarray}
which satisfies the current 95\% CL heavy neutrino mass limit, $m_{\nu_{4}} > 90.3$ GeV for Dirac coupling and $m_{\nu_{4}} > 80.5$ GeV for Majorana coupling~\cite{PDG}. Here the splitting of the fourth generation neutrino from the  three active neutrinos is caused by the hierarchical spectrum $M_{1}, M_{2}$ of right-handed Majorana fields, so it can be argued that the masses of $N^{(1)}_{R}$ and $N_{R5}$ have different origins, as shown in Eqs.~(\ref{A5Yukawa}) and~(\ref{MR}).  We note that $m_{\nu_{2}}$ is too small to accommodate the oscillation data. However, the three perturbations $\delta_{1,2,3}$ are different non-zero  quantities, and in general they can be varied at the same time independently. 
We find that the inclusion of a VEV for $H_{3}$ and the relaxation of the condition $v_{\phi_1} = v_{\phi_{2}} = v_{\phi_{3}}$ will alter the tribimaximal mixings and can lead to masses  in the range $10^{-1}-10^{-3}$ eV for the three light neutrinos for certain choices of parameters. \footnote{A thorough parameter space scan to obtain realistic neutrino mass spectrum is thus quite interesting but it is outside the scope of this work.} 
Therefore, we anticipate that higher order corrections will provide enough degrees of freedom to fit the experimental data and a degenerate mass spectrum of neutrino masses is preferred 
in the present model.

\subsection{Some Phenomenology}

\begin{figure}[ht]
  \centering
  \includegraphics[width=0.3\textwidth]{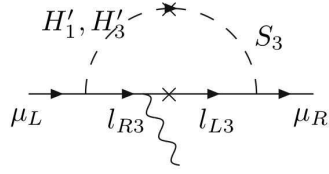}
  \caption{A 1-loop diagram for muon $(g-2)_\mu$ through new vector leptons. }
  \label{fig:g-2}
\end{figure}

The model contains many new sources of lepton flavor violation due to the extra scalars and fermions, and is similar to other flavor symmetry models. For example, lepton flavor violating processes in the $A_{4}$ models are studied in ref.~\cite{Ding:2009gh}. 

Here we address the presence of the extra contribution to the muon anomalous magnetic moment and leptonic rare decay processes ($\mu\rightarrow e\gamma, \tau \rightarrow \mu\gamma$) due to the existence of heavy vectorlike field $l_{L+R,3}$ which does not appear in other non-Abelian discrete symmetry models. The current limit of the muon anomalous $g-2$ is~\cite{PDG}
\begin{eqnarray}
\Delta a_{\mu} = (290\pm90)\times10^{-11} \;\;,
\end{eqnarray}
which is a $3.2\sigma$ deviation between SM calculations and experiment~\cite{Jegerlehner:2009ry}. The leading contribution to the muon anomalous magnetic moment in our model is showed in Fig.~\ref{fig:g-2}, and yields 
\begin{eqnarray}
\Delta a_{\mu} \approx \frac{Y_{4}Y_{S2}}{12\pi^2}\frac{m_{\mu}m_{L+R,_{3}}}{m^2_{H}}F(z) \;\;,
\end{eqnarray} 
where $z = \left(\frac{m_{L+R,3}}{m_{H}}\right)^{2}$. Note that we have ignored all the mixing factors in both Higgs 
and lepton sectors. 
The function $F(z)$ is defined as 
\begin{eqnarray}
F(z) = -\frac{3}{2(1 - z)^3}\left[3 - 4z + z^2 + 2\ln{z} \right] \;\;. 
\end{eqnarray}
Similar diagrams lead to rare radiative decays of $\mu \rightarrow e\gamma$ and $\tau \rightarrow \mu(e)\gamma$. The stringent bounds for these decays are~\cite{PDG} 
\begin{eqnarray}\label{LFVbound}
B(\mu\rightarrow e\gamma) < 1.2\times10^{-11}, \quad B(\tau\rightarrow e\gamma) < 1.1\times10^{-7}, \quad B(\tau\rightarrow\mu\gamma) < 4.5\times10^{-8}
\end{eqnarray}
at $90\%$ C.L. 
If the photon is on shell, we can write the invariant amplitude for the 
decays $l_{j} \rightarrow l_{i}\gamma$ in the form
\begin{eqnarray}
\emph{A}_{l_{j}l_{i}\gamma} = \frac{ie}{8\pi}\epsilon^*_{\mu}(q)\bar{u}_{i}(p-q)\left[\sigma^{\mu\nu}q_{\nu}(c_{1}P_{L} + c_{2}P_{R})\right]u_{j}(p) \;\;, 
\end{eqnarray}   
where the matrices $P_{L}$ and $P_{R}$ are defined by $P_{L} = (1 - \gamma_{5})/2$, $P_{R} = (1 + \gamma_{5})/2$ and $\sigma^{\mu\nu} = (i/2)[\gamma^{\mu},\gamma^{\nu}]$. The coefficients $c_{1}$ and $c_{2}$ were calculated in~\cite{He:2002pva} and are 
\begin{eqnarray}
c_{1} = \frac{Y_{4}Y_{S2}m_{\mu}}{\pi m^2_{H}}H(z) + \frac{Y^2_{S2}m_{L+R,3}}{\pi m^2_{H}}G(z) \quad {\rm and} \quad c_{2} = \frac{Y_{4}Y_{S2}m_{\mu}}{\pi m^2_{H}}H(z) + \frac{Y^2_{4}m_{L+R,3}}{\pi m^2_{H}}G(z)
\end{eqnarray} 
with 
\begin{eqnarray}
H(z) &=& -\frac{1}{6(1 - z)}\left[1 + \frac{3}{1 - z} - \frac{6}{(1 - z)^2} - \frac{6z}{(1 - z)^3\ln{z}}\right] \;\;, \\
G(z) &=& -\frac{1}{1 - z}\left[1 + \frac{2}{1 - z} + \frac{2}{(1 - z)^2\ln{z}}\right] \;\;.
\end{eqnarray}
The mixings between the charged leptons are ignored as before. If we take $m_{H} = 1$ TeV as the new physics scale and the limit of $m_{L+R,3} \approx m_{H}$, the anomalous muon magnetic moment and the branching ratio of $\mu\rightarrow e\gamma$ becomes 
\begin{eqnarray}
\Delta a_{\mu} \approx 5.1\times10^{-13}
\end{eqnarray}
and
\begin{eqnarray}
B(\mu\rightarrow e\gamma) \approx \frac{\alpha}{6\pi G^2_{F}m^2_{\mu}m^2_{H}}(Y^4_{S2} + Y^4_{4}) \approx 9.7\times10^{-11}
\end{eqnarray} 
respectively. Here we have used the relation $m_{\mu} = \sqrt{3}Y_{4}V'$ shown in Eq.~(\ref{chargemass}) and assumed the VEV of Higgs $V' \sim M_{W}$ and couplings $Y_{S2} = Y_{4}$. Similarly the estimate of the branching ratio for the $\tau\rightarrow\mu\gamma$ yields 
\begin{eqnarray}
B(\tau\rightarrow\mu\gamma) \approx 3.4\times10^{-13},
\end{eqnarray}
which is well below the current experimental bound given in Eq.~(\ref{LFVbound}).

\section{Discussion and conclusions}
We  minimally extend the three  generations  $A_{4}$  lepton flavor symmetry to a four generations $A_{5}$  lepton flavor symmetry. We find that the tribimaximal pattern in three generation neutrino mixings survives, and all the mass bounds on SM leptons can be satisfied. Notably, the electron  is predicted to be massless at tree level, hence one must calculate $m_{e}$ via quantum corrections. If the masses of the extra heavy leptons are within the reach of the LHC, we should be able to test the model. 
For a sequential fourth generation of leptons, the LHC with 
$\sqrt s = 7$ TeV and an integrated luminosity of 1 fb$^{-1}$ of data
can exclude fourth generation charged leptons with masses up to 250 GeV \cite{CRW}. 
It may be worth while to repeat this analysis for the $A_5$ model where
mixings between the sequential fourth generation and the extra vector-like leptons are allowed. 
The lepton flavor violating phenomenology in this model is very rich and needed to be studied further.
Before we close, we note that $A_{4}$ extended in the binary tetrahedral group $T'$ can provide a model of both the quark and lepton sectors \cite{Tprime}.
The three generation $T'$ model has calculable Cabibbo angle as well as other attractive features \cite{Tprime}. 
There is an analogous extension of the $A_{5}$ to the binary icosahedral group $I'$, where the attractive features of the $T'$ model can be utilized and four families of quarks and leptons can be accommodated simultaneously~\cite{CKY2}.

\acknowledgments
This work was supported in part by the US DOE grant DE-FG05-85ER40226, National Science Council of Taiwan under Grant No. 98-2112-M-001-014-MY3 
and the National Center for Theoretical Sciences of Taiwan (NCTS). CSC would like to thank L.~F. Li for useful discussion. TWK is grateful for the hospitality of the Physics Division of NCTS
where this work was initiated.

\appendix

\section{Discrete Symmetry Group $A_{5}$ and the Icosahedron}

\begin{figure}[ht]
  \centering
  \includegraphics[width=0.3\textwidth]{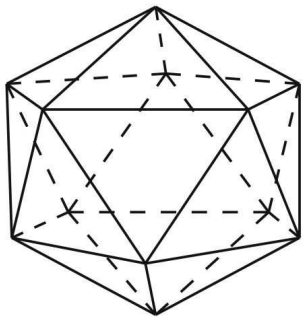}
  \caption{The regular icosahedron.}
  \label{fig:icosahedron}
\end{figure}

$A_{5}$ is a discrete symmetry group of even permutation of five objects. 
Its order, the number of elements, is equal to $(5!)/2=60$, which 
can be divided into 5 distinct conjugate classes. 
A regular icosahedron (see Fig.~\ref{fig:icosahedron}) consists of 20 equilateral triangles, 
30 edges and 12 vertices. The rotations of the icosahedron can be classified into five types, 
including $0$ or $2\pi$ rotation (the identity), $\pi$ rotations about the midpoint of each edge, 
$2\pi/5$ and $4\pi/5$ rotations about an axis through each vertex, 
and rotations by $2\pi/3$ about axes through the center of each face. 
These five types of rotations form five conjugate classes denoted by 
\begin{eqnarray}
C_{1},~C_{15},~C_{12},~C'_{12},~C_{20}.
\end{eqnarray}
Hence the icosahedral symmetry group $I$ also has 60 elements. 
One can show that the icosahedron symmetry group $I$ and the even permutation group $A_5$ are isomorphic.
The orthogonality relations for $A_{5}$ imply   $\sum^{5}_{i=1}n^2_{i} = 60$, 
where $n_i$ denotes the dimensionality of the irreducible representation $R_i$, and we know that the number of classes of a discrete group is equal to its number of irreducible representation.
Therefore, we have five irreducible representations (irreps): one trivial singlet $\textbf{1}$, two triplets $\textbf{3}$ and $\textbf{3}'$, one quartet $\textbf{4}$ and one quintet $\textbf{5}$. The character table and multiplication rules for  $A_{5}$  are shown in Table~{\ref{character}} and~\ref{multiplication} respectively. 

The sixty elements of $A_5$ can be generated by two elements, 
$s$ and $t$, which satisfy $s^2 = t^5 = (t^2st^3st^{-1}stst^{-1})^5 = e$. 
The explicit matrix representations of $s$ and $t$ for each irrep can be written down 
in terms of the golden ratio $\phi = \frac{1 + \sqrt{5}}{2}$~\cite{shirai}:
\begin{eqnarray}
&&\textbf{1}: s\rightarrow1, \quad t\rightarrow 1, \\
&&\textbf{3}: s=\frac{1}{2}\left(\begin{array}{ccc}-1 & \phi & \frac{1}{\phi} \\\phi & \frac{1}{\phi} & 1 \\\frac{1}{\phi} & 1 & -\phi\end{array}\right), \quad t=\frac{1}{2}\left(\begin{array}{ccc}1 & \phi & \frac{1}{\phi} \\-\phi & \frac{1}{\phi} & 1 \\\frac{1}{\phi} & -1 & \phi\end{array}\right), \\
&&\textbf{3}': s=\frac{1}{2}\left(\begin{array}{ccc}-\phi & \frac{1}{\phi} & 1 \\\frac{1}{\phi} & -1 & \phi \\1 & \phi & \frac{1}{\phi}\end{array}\right), \quad t=\frac{1}{2}\left(\begin{array}{ccc}-\phi & -\frac{1}{\phi} & 1 \\\frac{1}{\phi} & 1 & \phi \\-1 & \phi & -\frac{1}{\phi}\end{array}\right), \\
&&\textbf{4}: s=\frac{1}{4}\left(\begin{array}{cccc}-1 & -1 & -3 & -\sqrt{5} \\-1 & 3 & 1 & -\sqrt{5} \\-3 & 1 & -1 & \sqrt{5} \\-\sqrt{5} & -\sqrt{5} & \sqrt{5} & -1\end{array}\right), \quad t=\frac{1}{4}\left(\begin{array}{cccc}-1 & 1 & -3 & \sqrt{5} \\-1 & -3 & 1 & \sqrt{5} \\3 & 1 & 1 & \sqrt{5} \\\sqrt{5} & -\sqrt{5} & -\sqrt{5} & -1\end{array}\right), \\
&&\textbf{5}: s=\frac{1}{2}\left(\begin{array}{ccccc}\frac{1-3\phi}{4} & \frac{\phi^2}{2} & -\frac{1}{2\phi^2} & \frac{\sqrt{5}}{2} & \frac{\sqrt{3}}{4\phi} \\\frac{\phi^2}{2} & 1 & 1 & 0 & \frac{\sqrt{3}}{2\phi} \\-\frac{1}{2\phi^2} & 1 & 0 & -1 & -\frac{\sqrt{3}\phi}{2} \\\frac{\sqrt{5}}{2} & 0 & -1 & 1 & -\frac{\sqrt{3}}{2} \\\frac{\sqrt{3}}{4\phi} & \frac{\sqrt{3}}{2\phi} & -\frac{\sqrt{3}\phi}{2} & -\frac{\sqrt{3}}{2} & \frac{3\phi-1}{4}\end{array}\right), \quad t=\frac{1}{2}\left(\begin{array}{ccccc}\frac{1-3\phi}{4} & -\frac{\phi^2}{2} & -\frac{1}{2\phi^2} & -\frac{\sqrt{5}}{2} & \frac{\sqrt{3}}{4\phi} \\\frac{\phi^2}{2} & -1 & 1 & 0 & \frac{\sqrt{3}}{2\phi} \\\frac{1}{2\phi^2} & 1 & 0 & -1 & \frac{\sqrt{3}\phi}{2} \\-\frac{\sqrt{5}}{2} & 0 & 1 & 1 & \frac{\sqrt{3}}{2} \\\frac{\sqrt{3}}{4\phi} & -\frac{\sqrt{3}}{2\phi} & -\frac{\sqrt{3}\phi}{2} & \frac{\sqrt{3}}{2} & \frac{3\phi-1}{4}\end{array}\right).\nonumber \\
\end{eqnarray} 

\begin{center}
\begin{table}[ht]
\begin{tabular}{|c|c|c|c|c|c|}\hline
 & \textbf{1} & \textbf{3} & $\textbf{3}'$ & \textbf{4} & \textbf{5} \\ \hline 
$C_{1}$ & 1 & 3 & 3 & 4 & 5 \\ \hline 
$C_{15}$ & 1 & -1 & -1 & 0 & 1 \\ \hline 
$C_{20}$ & 1 & 0 & 0 & 1 & -1 \\ \hline 
$C_{12}$ & 1 & $\phi$ & $1 - \phi$ & -1 & 0 \\ \hline 
$C_{12}'$ & 1 & $1-\phi$ & $\phi$ & -1 & 0  \\\hline
\end{tabular}
\caption{\label{character} Character table of $A_{5}$ 
where $\phi = \frac{1 + \sqrt{5}}{2}$ is the golden ratio.}
\end{table}
\end{center} 

\begin{center}
\begin{table}[ht]
\begin{tabular}{|c|c|c|c|c|c|}\hline
$A_5$& \textbf{1} & \textbf{3} & $\textbf{3}'$ & \textbf{4} & \textbf{5} \\ \hline 
 \textbf{1} & \textbf{1} & \textbf{3} & $\textbf{3}'$ & \textbf{4} & \textbf{5} \\ \hline 
\textbf{3} & \textbf{3} & $\textbf{1}\oplus\textbf{3}\oplus\textbf{5}$  & $\textbf{4}\oplus\textbf{5}$ & $\textbf{3}'\oplus\textbf{4}\oplus\textbf{5}$ & $\textbf{3}\oplus\textbf{3}'\oplus\textbf{4}\oplus\textbf{5}$ \\ \hline 
$\textbf{3}'$ & $\textbf{3}'$ & $\textbf{4}\oplus\textbf{5}$ & $\textbf{1}\oplus\textbf{3}'\oplus\textbf{5}$ & $\textbf{3}\oplus\textbf{4}\oplus\textbf{5}$ & $\textbf{3}\oplus\textbf{3}'\oplus\textbf{4}\oplus\textbf{5}$ \\ \hline 
\textbf{4} & \textbf{4} & $\textbf{3}'\oplus\textbf{4}\oplus\textbf{5}$ & $\textbf{3}\oplus\textbf{4}\oplus\textbf{5}$ & $\textbf{1}\oplus\textbf{3}\oplus\textbf{3}'\oplus\textbf{4}\oplus\textbf{5}$ & $\textbf{3}\oplus\textbf{3}'\oplus\textbf{4}\oplus\textbf{5}\oplus\textbf{5}$ \\ \hline 
\textbf{5} & \textbf{5} & $\textbf{3}\oplus\textbf{3}'\oplus\textbf{4}\oplus\textbf{5}$ & $\textbf{3}\oplus\textbf{3}'\oplus\textbf{4}\oplus\textbf{5}$ & $\textbf{3}\oplus\textbf{3}'\oplus\textbf{4}\oplus\textbf{5}\oplus\textbf{5}$ & $\textbf{1}\oplus\textbf{3}\oplus\textbf{3}'\oplus\textbf{4}\oplus\textbf{4}\oplus\textbf{5}\oplus\textbf{5}$  \\\hline
\end{tabular}
\caption{\label{multiplication} Multiplication rules for the $A_{5}$ discrete group.}
\end{table}
\end{center}

\section{$A_{4}$ Multiplication Rules}
$A_{4}$ is a discrete symmetry corresponding to the even permutation of four objects. It has four irreducible representations: three in-equivalent one-dimensional representations $(\textbf{1}, \textbf{1}', \textbf{1}'')$ and a three-dimensional representation $\textbf{3}$. The multiplication rules are given 
in Table \ref{A4multiplication}.

\begin{center}
\begin{table}[ht]
\begin{tabular}{|c|c|c|c|c|}\hline
 $A_4$& \textbf{1} &$ \textbf{1}'$ & $\textbf{1}''$ & \textbf{3}   \\ \hline 
 \textbf{1} & \textbf{1} & $ \textbf{1}'$ & $ \textbf{1}''$& \textbf{3}   \\ \hline 
$ \textbf{1}'$ & $ \textbf{1}'$ & $ \textbf{1}''$  & $\textbf{1}$ & $\textbf{3}$ \\ \hline 
$ \textbf{1}''$& $ \textbf{1}''$ & $\textbf{3}$ & $ \textbf{1}'$ & $\textbf{3} $ \\ \hline 
\textbf{3} & \textbf{3} & $\textbf{3}$ & $\textbf{3}$ & $\textbf{1}\oplus  \textbf{1}' \oplus \textbf{1}''\oplus \textbf{3}\oplus \textbf{3} $  \\ \hline 
\end{tabular}
\caption{\label{A4multiplication} Multiplication rules for the $A_{4}$ discrete group.}
\end{table}
\end{center}

Given two triplets denoted by $a = (a_{1}, a_{2}, a_{3})$ and $b = (b_{1}, b_{2}, b_{3})$, 
the decomposition of $\textbf{3}\otimes \textbf{3} $ can be constructed explicitly as
\begin{eqnarray}
(a\otimes b)_{\textbf{1}} &=& a_{1}b_{1} + a_{2}b_{2} + a_{3}b_{3}\quad {,}  \nonumber \\
(a\otimes b)_{\textbf{1}'} &=& a_{1}b_{1} + \omega^2a_{2}b_{2} + \omega a_{3}b_{3}\quad {,}  \nonumber \\
(a\otimes b)_{\textbf{1}''} &=& a_{1}b_{1} + \omega a_{2}b_{2} + \omega^2a_{3}b_{3}\quad {,}  \\
(a\otimes b)_{\textbf{3}_{s}} &=& (a_{2}b_{3} + a_{3}b_{2}, a_{3}b_{1} + a_{1}b_{3}, a_{1}b_{2} + a_{2}b_{1}) \quad {,} \nonumber \\
(a\otimes b)_{\textbf{3}_{a}} &=& (a_{2}b_{3} - a_{3}b_{2}, a_{3}b_{1} - a_{1}b_{3}, a_{1}b_{2} - a_{2}b_{1}) \quad {,} \nonumber
\end{eqnarray}
where $\omega$ is the cube root of unity, $\omega = e^{i2\pi/3}$.

\section{Higgs potential}
In this Appendix we discuss the Higgs potential and its minimization in our model. The most general form of the Higgs potential containing the scalar fields $S_{4}$, $H_{4}$, $H'_{4}$ and $\Phi_{3}$, 
invariant under the discrete $A_{5}\times Z_{2}\times Z_{3}$ symmetries is given  by
\begin{eqnarray}
V &=& V(S_{4}) + V(H_{4}) + V(H'_{4}) + V(\Phi_{3}) + V(S_{4}, H_{4}) + V(S_{4}, H'_{4}) \nonumber \\
&& + V(S_{4}, \Phi_{3}) + V(H_{4}, H'_{4}) + V(H_{4}, \Phi_{3}) + V(H'_{4}, \Phi_{3}) + V(H_{4}, H'_{4}, \Phi_{3}) 
\end{eqnarray} 
where the individual terms are 
\begin{eqnarray}
V(S_{4}) &=& m^2_{s}S^2_{4} + \mu_{s}(S^2_{4})_{\textbf{4}}S_{4} + \lambda^{s}_{\alpha}(S^2_{4})_{\alpha}(S^2_{4})_{\alpha}, \\
V(H_{4}) &=& \mu^2_{H}(H^{\dag}_{4}H_{4})_{\textbf{1}} + \lambda^{H}_{\alpha}(H^{\dag}_{4}H_{4})_{\textbf{$\alpha$}}(H^{\dag}_{4}H_{4})_{\textbf{$\alpha$}}, \\ 
V(H'_{4}) &=& \mu^2_{H'}(H'^{\dag}_{4}H'_{4})_{\textbf{1}} + \lambda^{H'}_{\alpha}(H'^{\dag}_{4}H'_{4})_{\textbf{$\alpha$}}(H'^{\dag}_{4}H'_{4})_{\textbf{$\alpha$}}, \\ 
V(\Phi_{3}) &=& \mu^2_{\Phi}(\Phi^{\dag}_{3}\Phi_{3})_{\textbf{1}} + \lambda^{\Phi}_{\beta}(\Phi^{\dag}_{3}\Phi_{3})_{\textbf{$\beta$}}(\Phi^{\dag}_{3}\Phi_{3})_{\textbf{$\beta$}}, \\ 
V(S_{4}, H_{4}) &=& \delta^{HS}(H^{\dag}_{4}H_{4})_{\textbf{4}}S_{4} + \lambda^{HS}_{\alpha}(H^{\dag}_{4}H_{4})_{\textbf{$\alpha$}}(S^2_4)_{\textbf{$\alpha$}}, \\ 
V(S_{4}, H'_{4}) &=& \delta^{H'S}(H'^{\dag}_{4}H'_{4})_{\textbf{4}}S_{4} + \lambda^{H'S}_{\alpha}(H'^{\dag}_{4}H'_{4})_{\textbf{$\alpha$}}(S^2_4)_{\textbf{$\alpha$}}, \\  
V(S_{4}, \Phi_{3}) &=& \delta^{\Phi S}(\Phi^{\dag}_{3}\Phi_{3})_{\textbf{4}}S_{4} + \lambda^{\Phi S}_{\beta}(\Phi^{\dag}_{3}\Phi_{3})_{\textbf{$\beta$}}(S^2_{4})_{\textbf{$\beta$}}, \\ 
V(H_{4}, H'_{4}) &=& \lambda^{HH'}_{\alpha}(H^{\dag}_{4}H_{4})_{\textbf{$\alpha$}}(H'^{\dag}_{4}H'_{4})_{\textbf{$\alpha$}} + \lambda'^{HH'}_{\alpha}(H^{\dag}_{4}H'_{4})_{\textbf{$\alpha$}}(H'^{\dag}_{4}H_{4})_{\textbf{$\alpha$}} \nonumber \\
&& + \left[ \lambda''^{HH'}_{\alpha}(H^{\dag}_{4}H'_{4})_{\textbf{$\alpha$}}(H^{\dag}_{4}H'_{4})_{\textbf{$\alpha$}} + \rm{H.c.}\right], \\
V(H_{4}, \Phi_{3}) &=& \lambda^{H\Phi}_{\beta}(H^{\dag}_{4}H_{4})_{\textbf{$\beta$}}(\Phi^{\dag}_{3}\Phi_{3})_{\textbf{$\beta$}} + \lambda'^{H\Phi}_{\gamma}(H^{\dag}_{4}\Phi_{3})_{\textbf{$\gamma$}}(\Phi^{\dag}_{3}H_{4})_{\textbf{$\gamma$}} \nonumber \\
&& + \left[ \lambda''^{H\Phi}_{\gamma}(H^{\dag}_{4}\Phi_{3})_{\textbf{$\gamma$}}(H^{\dag}_{4}\Phi_{3})_{\textbf{$\gamma$}} + \rm{H.c.} \right],  \\
V(H'_{4}, \Phi_{3}) &=& \lambda^{H'\Phi}_{\beta}(H'^{\dag}_{4}H'_{4})_{\textbf{$\beta$}}(\Phi^{\dag}_{3}\Phi_{3})_{\textbf{$\beta$}} + \lambda'^{H'\Phi}_{\gamma}(H'^{\dag}_{4}\Phi_{3})_{\textbf{$\gamma$}}(\Phi^{\dag}_{3}H'_{4})_{\textbf{$\gamma$}} \nonumber \\
&& + \left[ \lambda''^{H'\Phi}_{\gamma}(H'^{\dag}_{4}\Phi_{3})_{\textbf{$\gamma$}}(H'^{\dag}_{4}\Phi_{3})_{\textbf{$\gamma$}} + \rm{H.c.} \right],  \\
V(H_{4}, H'_{4}, \Phi_{3}) &=& \lambda^{HH'\Phi}_{\gamma}(H^{\dag}_{4}\Phi_{3})_{\textbf{$\gamma$}}(H'^{\dag}_{4}H'_{4})_{\textbf{$\gamma$}} + \lambda'^{HH'\Phi}_{\gamma}(H'^{\dag}_{4}\Phi_{3})_{\textbf{$\gamma$}}(H'^{\dag}_{4}H_{4})_{\textbf{$\gamma$}} \nonumber \\
&& + \lambda''^{HH'\Phi}_{\gamma}(H^{\dag}_{4}\Phi_{3})_{\textbf{$\gamma$}}(H^{\dag}_{4}H_{4})_{\textbf{$\gamma$}} + \rm{H.c..}
\end{eqnarray}
Here we have introduced the notations $\alpha = \textbf{1}, \textbf{3}, \textbf{3}', \textbf{4}, \textbf{5}$; $\beta = \textbf{1}, \textbf{3}, \textbf{5}$; and $\gamma$ = $\textbf{3}', \textbf{4}, \textbf{5}$ respectively. $A_{5}$  is broken down to $A_4$ in the first stage of symmetry breaking, where it is not difficult to see that one can always choose the VEVs of the scalar field $S_{4}$ to the direction $\langle S_{4} \rangle = (V_{S}, 0, 0, 0)$, which can be the global minimum of $V(S_{4})$. The result of this allignment is that the $A_{5}$ breaking is responsible for the right-handed Majorana masses of $N_{R5}$ (Eq.~(\ref{MR})) and the remaining three components of   $S_{4}$, which form $S_{3}$ under $A_{4}$, will generate perturbations of active neutrino masses (Eq.~(\ref{MRp})-(\ref{numassp3})). As pointed out above, the scalar fields will decompose as $\textbf{4} \rightarrow \textbf{1} + \textbf{3}$ and $\textbf{3} \rightarrow \textbf{3}$ after $A_{5}$ breaks into $A_{4}$. One will obtain the $A_{4}$ symmetry potential with the collective coefficients $\delta's$ and $\lambda's$ of the decomposed fields $(S_{1}, S_{3}, H_{1}, H_{3}, H'_{1}, H'_{3}~\rm{and}~\Phi_{3})$ from Eqs.~(C2)-(C12). Now we turn to the minimization of the $A_{4}$ potential. 

\begin{figure}[ht]
  \centering
  \includegraphics[width=0.45\textwidth]{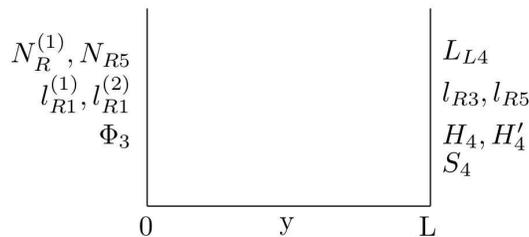}
  \caption{Fifth dimension and locations of the field configuration.}
  \label{fig:extrad}
\end{figure}

In our discussion all vacua can be accommodated by the large parameter space in the Higgs potential except the $\emph{ad hoc}$ vacuum alignment of $\Phi_{3}$ shown in Eq.~(\ref{vacuum}). The reason 
is the existence of the interaction terms between  the scalar fields $S_{4}, H_{4}, H'_{4}$ and $\Phi_{3}$ in $V(S_{4}, \Phi_{3}), V(H_{4}, \Phi_{3}), V(H'_{4}, \Phi_{3})$ and $V(H_{4}, H'_{4}, \Phi_{3})$. These terms will produce more independent equations derived from the extremum conditions than the unknown VEVs. The authors of Refs.~\cite{Altarelli:2005yp} and~\cite{He:2006dk}  showed how to deal with the vacuum alignment problem under the non-Abelian group symmetry. Here we  could extend the model with an extra spacial  dimension $y$, via the method  introduced in the first paper in~\cite{Altarelli:2005yp}. The fields are localized at the boundaries $y = 0$ and $y = L$ as shown in Fig.~\ref{fig:extrad} and the configurations will realize the needed vacuum. There are non-local effects involving both branes. A detailed explanation of this possibility is beyond the scope of this paper. 


\begin{thebibliography}{99}

\bibitem{He:2001tp}
  H.~J.~He, N.~Polonsky and S.~f.~Su,
  Phys.\ Rev.\  D {\bf 64}, 053004 (2001)
  [arXiv:hep-ph/0102144].

\bibitem{Erler-Langacker}
  J. Erler and P. Langacker, 
  Phys. Rev. Lett. 105, 031801(2010) [arXiv:1003.3211[hep-ph]].

\bibitem{Chanowitz}  
  M.~S.~Chanowitz,
  Phys.\ Rev.\  D {\bf 79}, 113008 (2009)
  [arXiv:0904.3570 [hep-ph]]; O.~Eberhardt, A.~Lenz and J.~Rohrwild,
  arXiv:1005.3505 [hep-ph].

\bibitem{PQ-Sher}  
  P.~Q.~Hung and M.~Sher,
  Phys.\ Rev.\  D {\bf 77}, 037302 (2008)
  [arXiv:0711.4353 [hep-ph]].

\bibitem{Higgs4th}  
  G.~D.~Kribs, T.~Plehn, M.~Spannowsky and T.~M.~P.~Tait,
  Phys.\ Rev.\  D {\bf 76}, 075016 (2007)
  [arXiv:0706.3718 [hep-ph]]; 
  T.~Cuhadar-Donszelmann, M.~Karagoz, V.~E.~Ozcan, S.~Sultansoy and  G.~Unel,
  JHEP {\bf 0810}, 074 (2008)
  [arXiv:0806.4003 [hep-ph]]; P.~Q.~Hung and C.~Xiong,
  arXiv:0911.3890 [hep-ph]; P.~Q.~Hung and C.~Xiong,
  arXiv:0911.3892 [hep-ph];  M.~S.~Chanowitz,
  Phys.\ Rev.\  D {\bf }, 035018 (2010)
  [arXiv:1007.0043 [hep-ph]]; S.~Dawson and P.~Jaiswal,
  arXiv:1009.1099 [hep-ph].
      
\bibitem{LHCb} 
  W.~S.~Hou, M.~Nagashima and A.~Soddu,
  Phys.\ Rev.\  D {\bf 76}, 016004 (2007)
  [arXiv:hep-ph/0610385]; W.~S.~Hou, H.~n.~Li, S.~Mishima and M.~Nagashima,
  Phys.\ Rev.\ Lett.\  {\bf 98}, 131801 (2007)
  [arXiv:hep-ph/0611107];  A.~Soni, A.~K.~Alok, A.~Giri, R.~Mohanta and S.~Nandi,
  Phys.\ Lett.\  B {\bf 683}, 302 (2010)
  [arXiv:0807.1971 [hep-ph]]; M.~Bobrowski, A.~Lenz, J.~Riedl and J.~Rohrwild,
  Phys.\ Rev.\  D {\bf 79}, 113006 (2009)
  [arXiv:0902.4883 [hep-ph]]; W.~S.~Hou, Y.~Y.~Mao and C.~H.~Shen,
  Phys.\ Rev.\  D {\bf 82}, 036005 (2010)
  [arXiv:1003.4361 [hep-ph]].
  
\bibitem{Holdom}
  B.~Holdom,
  JHEP {\bf 0703}, 063 (2007)
  [arXiv:hep-ph/0702037], 
  JHEP {\bf 0708}, 069 (2007)
  [arXiv:0705.1736 [hep-ph]]. 
  
\bibitem{CRW}
  L. M. Carpenter, A. Rajaraman and D. Whiteson, 
  arXiv:1010.1011 [hep-ph]; A.~Lenz, H.~Pas and D.~Schalla,
  arXiv:1010.3883 [hep-ph].
  
\bibitem{overview}
  B.~Holdom, W.~S.~Hou, T.~Hurth, M.~L.~Mangano, S.~Sultansoy and G.~Unel,
  PMC Phys.\  A {\bf 3}, 4 (2009)
  [arXiv:0904.4698 [hep-ph]].
  
\bibitem{Ma:2001dn} 
  E.~Ma and G.~Rajasekaran, Phys.\ Rev.\  D {\bf 64}, 113012 (2001) [arXiv:hep-ph/0106291]; 
  K.~S.~Babu, E.~Ma and J.~W.~F.~Valle,
  Phys.\ Lett.\  B {\bf 552}, 207 (2003)
  [arXiv:hep-ph/0206292]; E.~Ma,
  Phys.\ Rev.\  D {\bf 70}, 031901 (2004)
  [arXiv:hep-ph/0404199]; E.~Ma,
  Phys.\ Rev.\  D {\bf 72}, 037301 (2005)
  [arXiv:hep-ph/0505209]; E.~Ma,
  Mod.\ Phys.\ Lett.\  A {\bf 21}, 2931 (2006)
  [arXiv:hep-ph/0607190]. 
  
\bibitem{Altarelli:2005yp}   
  G.~Altarelli and F.~Feruglio,
  Nucl.\ Phys.\  B {\bf 720}, 64 (2005)
  [arXiv:hep-ph/0504165]; G.~Altarelli and F.~Feruglio,
  Nucl.\ Phys.\  B {\bf 741}, 215 (2006)
  [arXiv:hep-ph/0512103]; G.~Altarelli, F.~Feruglio and C.~Hagedorn,
  JHEP {\bf 0803}, 052 (2008)
  [arXiv:0802.0090 [hep-ph]];
  
 \bibitem{Zee:2005ut} 
  A.~Zee,
  Phys.\ Lett.\  B {\bf 630}, 58 (2005)
  [arXiv:hep-ph/0508278]; 
  
 \bibitem{He:2006dk} 
  X.~G.~He, Y.~Y.~Keum and R.~R.~Volkas, 
  JHEP {\bf 0604}, 039 (2006) [arXiv:hep-ph/0601001]; 
  
\bibitem{King:2006np} 
  S.~F.~King and M.~Malinsky,
  Phys.\ Lett.\  B {\bf 645}, 351 (2007)
  [arXiv:hep-ph/0610250]; 

\bibitem{Morisi:2007ft}  
  S.~Morisi, M.~Picariello and E.~Torrente-Lujan,
  Phys.\ Rev.\  D {\bf 75}, 075015 (2007)
  [arXiv:hep-ph/0702034];  
 
\bibitem{Bazzocchi:2007na}  
  F.~Bazzocchi, S.~Kaneko and S.~Morisi,
  JHEP {\bf 0803}, 063 (2008)
  [arXiv:0707.3032 [hep-ph]];
  
\bibitem{Frampton:2008ci}  
   P.~H.~Frampton and S.~Matsuzaki,
  arXiv:0806.4592 [hep-ph].

\bibitem{Datta:2003qg}
  A.~Datta, F.~S.~Ling and P.~Ramond,
  Nucl.\ Phys.\  B {\bf 671}, 383 (2003)
  [arXiv:hep-ph/0306002].
  
\bibitem{Kajiyama:2007gx}
  Y.~Kajiyama, M.~Raidal and A.~Strumia,
  Phys.\ Rev.\  D {\bf 76}, 117301 (2007)
  [arXiv:0705.4559 [hep-ph]].

\bibitem{ratio} Euclid of Alexandria, Elements, Book 6, Definition 3.---{\it A straight line 
is said to have been cut in extreme and mean ratio when, as the whole line is to the 
greater segment, so is the greater to the less.}  

\bibitem{Cummins:1988dr}
  C.~J.~Cummins and J.~Patera,
  J.\ Math.\ Phys.\  {\bf 29} (1988) 1736.
  
\bibitem{shirai} 
  K.~Shirai, J.\ Phys.\ Soc.\ Jpn. {\bf 61}, 2735 (1992).   

\bibitem{Luhn:2007yr}
  C.~Luhn, S.~Nasri and P.~Ramond,
  J.\ Math.\ Phys.\  {\bf 48}, 123519 (2007)
  [arXiv:0709.1447 [hep-th]].
  
\bibitem{Ludl:2009ft}
  P.~O.~Ludl,
  arXiv:0907.5587 [hep-ph].
  
\bibitem{Ishimori:2010au}
  H.~Ishimori, T.~Kobayashi, H.~Ohki, H.~Okada, Y.~Shimizu and M.~Tanimoto,
  Prog.\ Theor.\ Phys.\ Suppl.\  {\bf 183}, 1 (2010)
  [arXiv:1003.3552 [hep-th]].

\bibitem{Everett:2008et}
  L.~L.~Everett and A.~J.~Stuart,
  Phys.\ Rev.\  D {\bf 79}, 085005 (2009)
  [arXiv:0812.1057 [hep-ph]].  
  
\bibitem{Rodejohann:2008ir}
  W.~Rodejohann,
  Phys.\ Lett.\  B {\bf 671}, 267 (2009)
  [arXiv:0810.5239 [hep-ph]].  
  
\bibitem{Barr:1976bk}
  S.~M.~Barr and A.~Zee,
  Phys.\ Rev.\  D {\bf 15}, 2652 (1977).
    
\bibitem{Achard:2001qw}
  P.~Achard {\it et al.}  [L3 Collaboration],
  Phys.\ Lett.\  B {\bf 517}, 75 (2001)
  [arXiv:hep-ex/0107015].
  
\bibitem{Georgi:1972hy}
  H.~Georgi and S.~L.~Glashow,
  Phys.\ Rev.\  D {\bf 7}, 2457 (1973).

\bibitem{Barr:1978rv}
  S.~M.~Barr and A.~Zee,
  Phys.\ Rev.\  D {\bf 17}, 1854 (1978).

\bibitem{Balakrishna:1988xg}
  B.~S.~Balakrishna,
  Phys.\ Lett.\  B {\bf 214}, 267 (1988).
  
\bibitem{Ibanez:1982xg}
  S.~M.~Barr,
  Phys.\ Rev.\  D {\bf 21}, 1424 (1980); L.~E.~Ibanez,
  Phys.\ Lett.\  B {\bf 117}, 403 (1982); B.~S.~Balakrishna,
  Phys.\ Rev.\ Lett.\  {\bf 60}, 1602 (1988); K.~S.~Babu and E.~Ma,
  Mod.\ Phys.\ Lett.\  A {\bf 4}, 1975 (1989); X.~G.~He, R.~R.~Volkas and D.~D.~Wu,
  Phys.\ Rev.\  D {\bf 41}, 1630 (1990); K.~S.~Babu and R.~N.~Mohapatra,
  Phys.\ Rev.\ Lett.\  {\bf 64}, 2747 (1990).  

\bibitem{Barr:2007ma}   
  S.~M.~Barr,
  Phys.\ Rev.\  D {\bf 76}, 105024 (2007)
  [arXiv:0706.1490 [hep-ph]]; B.~A.~Dobrescu and P.~J.~Fox,
  JHEP {\bf 0808}, 100 (2008)
  [arXiv:0805.0822 [hep-ph]].
 
\bibitem{Ma:2004zv}
  E.~Ma,
  Phys.\ Rev.\  D {\bf 70}, 031901 (2004)
  [arXiv:hep-ph/0404199].  
  
\bibitem{Ding:2009gh}
  F.~Feruglio, C.~Hagedorn, Y.~Lin and L.~Merlo,
  Nucl.\ Phys.\  B {\bf 809} (2009) 218
  [arXiv:0807.3160 [hep-ph]]; C.~Hagedorn, E.~Molinaro and S.~T.~Petcov,
  JHEP {\bf 1002} (2010) 047
  [arXiv:0911.3605 [hep-ph]]; F.~Feruglio, C.~Hagedorn, Y.~Lin and L.~Merlo,
  arXiv:0911.3874 [hep-ph]; G.~J.~Ding and J.~F.~Liu,
  JHEP {\bf 1005}, 029 (2010)
  [arXiv:0911.4799 [hep-ph]].
  

\bibitem{PDG} 
  C. Amsler et al. (Particle Data Group), Phys. Lett. B {\bf 667}, 1 (2008) and 
  2009 partial update for the 2010 edition.


\bibitem{Jegerlehner:2009ry}
  F.~Jegerlehner and A.~Nyffeler,
  Phys.\ Rept.\  {\bf 477}, 1 (2009)
  [arXiv:0902.3360 [hep-ph]].
  
\bibitem{He:2002pva}
  B.~He, T.~P.~Cheng and L.~F.~Li,
  Phys.\ Lett.\  B {\bf 553}, 277 (2003)
  [arXiv:hep-ph/0209175]; P.~Q.~Hung,
  Phys.\ Lett.\  B {\bf 659}, 585 (2008)
  [arXiv:0711.0733 [hep-ph]]. 

 \bibitem{Tprime}  

  P.~H.~Frampton and T.~W.~Kephart,
  Int.\ J.\ Mod.\ Phys.\  A {\bf 10}, 4689 (1995)
  [arXiv:hep-ph/9409330];
  A.~Aranda, C.~D.~Carone and R.~F.~Lebed,
  Phys.\ Rev.\  D {\bf 62}, 016009 (2000)
  [arXiv:hep-ph/0002044];
  M.~C.~Chen and K.~T.~Mahanthappa,
  Phys.\ Lett.\  B {\bf 652}, 34 (2007)
  [arXiv:0705.0714 [hep-ph]];
   P.~H.~Frampton and T.~W.~Kephart,
  JHEP {\bf 0709}, 110 (2007)
  [arXiv:0706.1186 [hep-ph]];
   P.~H.~Frampton, T.~W.~Kephart and S.~Matsuzaki,
  Phys.\ Rev.\  D {\bf 78}, 073004 (2008)
  [arXiv:0807.4713 [hep-ph]];
  P.~H.~Frampton and S.~Matsuzaki,
  Phys.\ Lett.\  B {\bf 679}, 347 (2009)
  [arXiv:0902.1140 [hep-ph]].
 


\bibitem{CKY2} 
  C.-S. Chen, T.~W.~Kephart and T.~C. Yuan, work in progress.

\end{thebibliography}
\end{document}